\documentclass[11pt]{article}
\usepackage[brazil]{babel}
\usepackage{url}
\usepackage{amssymb,amsxtra}
\usepackage{amsmath,color,graphicx}
\usepackage{lineno}
\usepackage{graphicx}
\RequirePackage{geometry}
\usepackage{indentfirst}

\begin{document}

\def\chaptername{}
\def\contentsname{Sum\'{a}rio}
\def\listfigurename{Figuras}
\def\listtablename{Tabelas}
\def\abstractname{Resumo}
\def\appendixname{Ap\^{e}ndice}
\def\refname{\large Refer\^{e}ncias bibliogr\'{a}ficas}
\def\bibname{Bibliografia}
\def\indexname{\'{I}ndice remissivo}
\def\figurename{\small Fig.~}
\def\tablename{\small Tab.~}
\def\pagename{\small Pag.}
\def\seename{veja}
\def\alsoname{veja tamb\'em}
\def\na{-\kern-.4em\raise.8ex\hbox{{\tt \scriptsize a}}\ }
\def\pa{\slash \kern-.5em\raise.1ex\hbox{p}\ }
\def\ro{-\kern-.4em\raise.8ex\hbox{{\tt \scriptsize o}}\ }
\def\no{n$^{\underline{\rm o}}$}

\setcounter{tocdepth}{3}

\clearpage
\pagenumbering{arabic}

\thispagestyle{empty}
\parskip 8pt

\vspace*{0.2cm}
\begin{center}

{\huge \bf  A quarta dimens\~{a}o: da sua natureza espacial na geometria euclidiana \`{a} componente tipo tempo de variedades n\~{a}o euclidianas}\\

\ \\

{\huge \bf  The Fourth Dimension: From its spatial nature in Euclidean geometry to a time-like component of non-Euclidean manifolds}\\

\vspace*{1.0cm}
{\Large \bf \it Jos\'{e} Maria Filardo Bassalo;$^{1}$ Francisco Caruso;$^{2,3}$ Vitor Oguri$^3$}\\[2.em]

{{$^{1}$ Academia Paraense de Ci\^{e}ncias \& Funda\c{c}\~{a}o Minerva, Bel\'{e}m, Par\'{a}, Brasil.}}

{{$^{2}$ Centro Brasileiro de Pesquisas F\'{\i}sicas, Coordena\c{c}\~{a}o de F\'{\i}sica de Altas Energias, 22290-180, Rio de Janeiro, RJ, Brasil.}}

{{$^{3}$ Universidade do Estado do Rio de Janeiro, Instituto de F\'{\i}sica Armando Dias Tavares, 20550-900, Rio de Janeiro, RJ, Brasil.}}
\vfill
\end{center}

\noindent \textbf{Resumo}

Neste artigo, apresenta-se a evolu\c{c}\~{a}o das ideias sobre a quarta dimens\~{a}o espacial, partindo daquelas que surgem da geometria euclidiana cl\'{a}ssica e abordando, em seguida, as que resultam do \^{a}mbito das geometrias n\~{a}o euclidianas, como as de Riemann e Minkowski. Particular aten\c{c}\~{a}o \'{e} dada ao momento no qual o tempo real passa efetivamente a ser considerado como uma quarta dimens\~{a}o, conforme introduzido por Einstein.

\noindent \textbf{Palavras-chave:} Quarta Dimens\~{a}o; Espa\c{c}o; Geometrias n\~{a}o Euclidianas; Hist\'{o}ria da Geometria; Hist\'{o}ria da F\'{\i}sica.

\vspace*{0.7cm}

\noindent \textbf{Abstract}

In this article, the evolution of the ideas about the fourth spatial dimension is presented, starting from those which come out within classical Euclidean geometry and going through those arose in the framework of non-Euclidean geometries, like those of Riemann and Minkowski. Particular attention is given to the moment when real time is effectively considered as a fourth dimension, as introduced by Einstein.


\vfill

\newpage

\section{The beginning of the history: not more than three}\label{beginning}
\label{intro}
The first step in the history of the fourth dimension was actually an attempt to deny its existence. Indeed, the impossibility of a fourth dimension was actually sustained by Aristotle of Stagira (384-322 b.C.). In effect, in his \textit{De Caelo}, which consists of Four Books, he treated this impossibility just right in the first paragraph of Book 1, saying, in summary, that:

\begin{quotation}
\noindent\baselineskip=10pt {\small
\noindent ``A magnitude if divisible one way is a line, if two ways a surface, and if three a body. Beyond these there is no other magnitude, because the three dimensions are all that there are, and that which is divisible in three directions is divisible in all''~\cite{Barnes}.}
\end{quotation}

In this same paragraph, the Stagirite continues giving a cosmological justification of this number three by appealing to its divinization, sustained by the Pythagoreans. Quoting him,

\begin{quotation}
\noindent\baselineskip=10pt {\small
\noindent ``For, as the Pythagoreans say, the universe and all that is in it is determined by the number three, since beginning and middle and end give the number of the universe, and the number they give is a triad. And so, having taken these three from nature as (so to speak) laws of it, we make further use of the number three in the worship of the Gods''.}
\end{quotation}

Such kind of identification between the tri-dimensionality of space and God's will is recurrent in the history of science. Johannes Kepler (1571-1630), for example, asseverated that three is exactly the number of dimensions due to the Holly Trinity~\cite{Pauli,Jung}.

The second necessary (but not sufficient) step toward the conception of the fourth dimension has to do with the systematization of the geometric knowledge in Ancient Greece.

The Greek mathematician Euclid of Alexandria (c.~323-285 b.C.) wrote his famous treatise on Geometry, \textit{The Elements}, which has influenced the development of Western Mathematics for more than 2\,000 years. This classical work contains Thirteen Books. From Book 1 to 10 just Plane Geometry is considered. In Book 1, Euclid presents 23 definitions including those of point, line and surface, as following: 1) ``A point is that which has no part''; 2) ``A line is breadthless length''; (...); 5) ``A surface is that which has length and breadth only''~\cite{Euclid-01}.

The third dimension is treated only in the last three Books, when a solid is defined in the following way: ``A solid is that which has length, breadth and depth''~\cite{Euclid-02}.

Throughout the entire compendium, Euclid limits himself to treat Linear, Plane and Spatial Geometry, and he did not even consider the possibility of a fourth dimension.

Some centuries later, the Greek astronomer Claudius Ptolemy (85-165), in his (lost) book \textit{On Distance}, published in 150~a.C., gave a ``proof'' about the impossibility of the fourth dimension, based on the very fact that it is impossible to draw a fourth line perpendicular to three mutually perpendicular lines. This is indeed not a proof, but rather reinforce that we are not able to visualize the fourth dimension from which one cannot conclude about its non existence.

To the best of our knowledge, speculations and new ideas about the existence of a fourth dimension had to wait for the middle of the 16th century on to be strengthened, when a more propitious intellectual atmosphere is to be found, as we will see all over this paper.

For the moment, it is important to stress that the two greater synthesis of the Classical Greek Philosophy -- that of Aristotle and that of Euclid -- considered impossible the existence of more than three spatial dimensions. This very fact is not meaningless, so far the enormous influence of these two thinkers is considered. The long period in which there was no discussion why space could not have a different dimensionality may be, in part, attributed to Aristotle's authority and, on the other side, to the fact that the study of Euclid's \textit{Elements} in the Middle Ages, including different educational levels, was languished or quite neglected~\cite{Heath}.

The overcoming of the domination of the Aristotelian influence and the abandonment of its Cosmos must still await the Renaissance~\cite{Koyre-1,Livro}. As Koyr\'{e} emphasizes, this implies and imposes the reformulation of the basic principles of philosophical and scientific reason~\cite{Koyre-2}. In contrast, the revival of interest in Euclid's \textit{Elements} should wait the invention of printing press by Johannes Gutenberg (c.~1400-1468)~\cite{McMurtie}.

It is not out of place to remember that, since the first prehistoric cave painting until the medieval period, the World was pictured without any perspective in two dimensional spaces, probably translating the difficulty to represent the third dimension on a bi-dimensional canvas, wall or any other surface. To go further required a good understanding and domain of Geometry.

The geometrization of space and the desire to represent it in painting had an enormous impact on the Italian Art in the end of the \textit{Trecento} and during the following centuries. Actually, the first to introduce the tridimensionality in Medieval Art~\cite{Bunim} was the Italian painter Giotto di Bondone (1266-1337). Giotto painted almost all the walls of St.~Francis' Lower Church, in Assisi. ``The Legend of St.~Francis'', whose authorship is traditionally attributed to him, is the theme of a cycle of 28 frescoes that are found in the Upper Basilica in Assisi, probably painted between 1297 and 1300. The frescoes painted by him in the Arena Chapel at Pauda, about the year 1305, mark an entirely new stage in the development of empirical perspective~\cite{White}. He was also the first artist in that period to paint the Sky in blue, diverting up from the traditional golden Sky characteristic of the Byzantine Art~\cite{Caruso-01,Caruso-arte}. Clearly, he was inspired by St.~Francis' world view, who pointed out emphatically that there was a second book (not only the Sacred Scripture) able to bring someone to God: it is the \textit{Book of Nature}. One should look at Nature as it is. This attitude suggests and anticipates a transition to the new relation between Man and Nature, which is a prelude to a new man that is still to be forged in the Renascence~\cite{Cassirer}.

The formal discovery of perspective is attributed to the Italian architect Filippo Brunelleschi (1377-1446), who suggested a system that explained how objects shrink in size according to their position and distance from the eye. In 1435, in a book named \textit{On Painting}, Leon Battista Alberti (1404-1472) provided the first theory of what we now call linear perspective.
The tri-dimensional representations of painting conquered then a scientific aspect when painters and architects of the \textit{Quattrocento} started to study the relationship between Geometric Optics~\cite{Light} and Perspective in the Euclidean space, as did, for instance, Piero della Francesca~\cite{Francesca}.

\section{The first ideas of a new dimension in space}
\label{sec:2}

Back to the fourth dimension, the idea of a new spatial dimension was revived by the studies of several mathematicians in the 16th and 17th centuries. Indeed, the Italian physicist, philosopher, mathematician and physician Ge(i)rolamo Cardano (1501-1576) and the French mathematician Fran\c{c}ois Vi\`{e}te (1540-1603) considered such ``additional'' dimension in their researches on quadratic and cubic equations. The same did the French mathematician and physicist Blaise Pascal (1623-1662) in his study named \textit{Trait\'{e} des trilignes rectangles et le leurs onglets}~\cite{Pascal}, when, generalizing his ``trilignes'' from the plane to the space and beyond, he wrote: ``\textit{The fourth dimension is not against the pure Geometry}.''

Meanwhile, the French philosopher and mathematician Ren\'{e} du Perron Descartes (1596-1650), as is well known -- and following the same pragmatic view of Aristotle, Euclid, and others concerning space (Section~\ref{beginning}) --, holds that ``\textit{the nature of matter, or body consists (...) simply in its being something which is extended in length, breadth, and depth}''~\cite{Descartes}.

However, in his \textit{Treatise on Algebra}, published in 1685, the English mathematician John Wallis (1616-1703) condemned again the existence of a higher-dimensional space saying that it is a ``\textit{monster in nature, less possible than a chimera or a centaur}''~\cite{Wallis}. And continues: ``\textit{Length, Breadth and Thickness take up the whole of space. Nor can Fansie imagine how there should be a Fourth Local Dimension beyond these Three}''~\cite{Boyer},~\cite{Struik},~\cite{Kline}.

The young Immanuel Kant (1724-1804), in his doctoral thesis (1747), tried to explain why is space three dimensional~\cite{Kant-tese}. Actually, he did not succeed~\cite{Kant-Studien}, but this work has two important merits: Kant pointed out that space dimensionality should be understood in the framework of Physics, which proved to be a fruitful idea in the 20th century~\cite{fundamenta}, and he concluded his speculations by imagining various types of spaces -- which came true later in the 19th century -- and alluding to them with these words of hope: ``\textit{A science of all these possible kinds of space would undoubtedly be the highest enterprise which a finite understanding could undertake in the field of geometry.}''~\cite{Kant-citation}.
%``Eine Wissenschaft von allen diesen m\"{o}glichen Raumesarten w\"{a}re unfehlbar die h\"{o}chste Geometrie, die ein endlicher Verstand unternehmen k\"{o}nnte''.

During the 18th century, the theme of the fourth dimension was treated again from a different perspective, \textit{i.e.}, by associating it to \textit{time} no more to \textit{space}. We are talking about the contribution of the French mathematician Jean le Rond d'Alembert (1717-1783) and his proposal in the entry ``Dimension'' wrote for the \textit{Encyclop\'{e}die ou Dictionnaire Raisonn\'{e} des Sciences, des Arts, et de M\'{e}tiers}, published between 1751 and 1772, by Denis Diderot (1713-1784) and himself.

Time was considered also as a fourth dimension by the Italian-French mathematician and astronomer Joseph-Louis Lagrange (1736-1813), in his books \textit{M\'{e}canique Analytique}, de 1788, and \textit{Th\'{e}orie des Fonctions Analytiques}, de 1797. Later, Lagrange says something like: One can consider the Mechanics as a Geometry in four dimensions and the Analytical Mechanics as an extension of the Analytical Geometry, developed by Descartes in his book \textit{La G\'{e}om\'{e}trie}, published in 1637~\cite{Kline}.

In the beginning of the 19th century, more specifically in 1827, in the book \textit{Der Barycentrische Calcul}, the German mathematician Augustus Ferdinand M\"{o}bius (1790-1868) rejected the existence of the fourth dimension when he observed that geometrical figures cannot be superimposed in three dimensions since they are the mirror images of themselves~\cite{Moebius}. Such a superposition, however, could happen just in a four dimensional space but, ``\textit{since, however, such a space cannot be thought about, the superposition is impossible}''~\cite{Kline}.

The fourth dimension was also proposed by the German physicist and mathematician Julius Pl\"{u}cker (1801-1868) in his book entitled \textit{System der Geometrie des Raumes}, published in 1846, in which he affirm that planes are nothing but collections of lines, as the intersection of them results in points. Following this idea, Pl\"{u}cker said that if lines are fundamental elements of space, then space is four-dimensional, because it is necessary four parameters to cover all the space with lines.  However, this proposal was rejected because it was saw as metaphysics. But, in any case, it was quite clear for many mathematicians that the three-dimensional Geometry had to be generalized~\cite{Manning}.

It is important to stress that before, in 1748, and later, in 1826, the Swiss physicist and mathematician Leonhard Euler (1707-1783) and the French mathematician Augustine Louis Cauchy (1789-1857), respectively, had tried to represent lines in space. In 1843, the English mathematician Arthur Cayley (1821-1895) had developed the Analytical Geometry in a $n$-dimensional space, taking the theory of determinants (name due to Cauchy) as a tool. Soon, in 1844, the German mathematician Hermann G\"{u}nter Grassmann (1809-1877) published the book \textit{Die Lineale Ausdehnungslehre, ein neuer Zweig der Mathematik}, in which he thought on a $n$-dimensional Geometry, stimulated by the discovery of the quaternion, announced by the Irish mathematician and physicist Sir William Rowan Hamilton (1805-1865), in 1843~\cite{Boyer,Kline}.

Actually, the conjectures about the fourth dimension acquire more soundness from the development of the so-called non-Euclidean Geometries in the 19th century~\cite{Caruso-02}. Let us now summarize how it happened.

\section{The new background of non-Euclidean geometries}\label{back}
\label{sec:3}

It is attributed to the Greek philosopher and geometer Thales of Miletus (c. 624 - c. 546) the demonstration of the following theorems: In isosceles triangles, the angles at the base are equal to one another, and, if the equal straight lines be produced further, the angles under the base will be equal to one another~\cite{Euclid-03}; If two straight lines cut one another, they make the vertical angles equal to one another~\cite{Euclid-04};  Those theorems allow one to prove the so-called Thales Theorem: (...), and the three interior angles of the triangle are equal to two right angles~\cite{Euclid-05}.  This theorem was considered as a divine truth by the influent Italian philosopher and theologian Thomas Aquinas (1225-1274), when, in his famous \textit{Summa Theologica}, issued around 1265, sustained to have proved that God could not construct a triangle for which the internal angles summed up more than 180$^{\circ}$. It is opportune to remember that Thales Theorem is also a consequence of the famous Postulate 5 of Euclid Book 1: That, if a straight line falling on two straight lines make the interior angles on the same side less than two right angles, the two straight lines, if produced indefinitely, meet on that side on which are the angles less than the two right angles~\cite{Euclid-06}.

In 1795, this Postulate number 5 was enounced by the English mathematician John Playfair (1748-1819) as follow: Through a given point only one parallel can be drawn to a given straight line. This is known as the Parallel Postulate~\cite{Euclid-07}.

The Parallel Postulate started to be criticized by the German mathematician and physicist Johann Carl Friedrich Gauss (1777-1855) -- who invented the concept of curvature --, in the last decade of the 18th century, when he tried to demonstrate it by using Euclidean Geometry. In effect, in 1792, when he was fifteen years old, he wrote a letter to his friend the German astronomer Heinrich Christian Schumacher (1780-1850), in which he discussed the possibility of having a Logical Geometry where the Parallel Postulate did not hold. In 1794, he conceived a new Geometry for which the area of a quadrangular figure should be proportional to the difference between $360^\circ$ and the sum of its internal angles. Later, in 1799, Gauss wrote a letter to his friend and Hungarian mathematician Wolfgang Farkas Bolyai (1775-1856) saying that he had tried, without success, to deduce the Parallel Postulate from other postulates of Euclidean Geometry~\cite{Kaku}.

During the 19th century, Gauss continued the discussion with friends on the plausibility of the existence of a Non-Euclidean Geometry. So, around 1813, he developed what he initially called Anti-Euclidean Geometry, then Astral Geometry and, finally, Non-Euclidean Geometry. He was so convinced about the existence of this new Geometry that he wrote a letter, in 1817, to his friend and German astronomer and physician Heinrich Wilhelm Matth\"{a}us Olbers (1758-1840), stressing the physical necessity of such a Geometry as follow~\cite{Kline-02}:

\begin{quotation}
\noindent\baselineskip=10pt {\small
\noindent ``I am becoming more and more convinced that the [physical] necessity of our [Euclidean] geometry cannot be proved, at least not by human reason nor for human reason. Perhaps in another life we will be able to obtain insight into the nature of space, which is now unattainable. Until then we must place geometry not in the same class with arithmetic, which is purely a priori, but with mechanics.''}
\end{quotation}

Seven years later, in 1824, answering a letter from the German mathematician Franz Adolf Taurinus (1794-1874) talking about a demonstration he did that the sum of the internal angles of a triangle cannot be neither greater nor smaller than 180$^{\circ}$, Gauss told him that there was not geometrical rigor in that demonstration because, in spite of the fact that the ``metaphysicists'' consider the Euclidean Geometry as the truth, this Geometry is incomplete. The ``metaphysicists'' quoted by Gauss were the followers of Kant, who wrote, in 1781, in his \textit{Kritik der reinen Vernunft}~\cite{Kant-02}, more precisely in its first chapter entitled Transcendental Doctrine of Elements what follows: a) Space is not a conception which has been derived from outward experiences; b) Space then is a necessary representation \textit{a priori}, which serves for the foundation of all external intuitions; c) Space is represented as an infinite given quantity; d) Space has only three dimensions; d) (...)  possibility of geometry, as a synthetic science \textit{a priori}, becomes comprehensible~\cite{Kant-03}.

Although we owe to Gauss the discovery of Non-Euclidean Geometry, he did not have the courage to publish his discoveries. Indeed, in a letter sent to a German friend and astronomer Friedrich Wilhelm Bessel (1784-1846), in 1829, Gauss affirm that he probably would never publish his findings in this subject because he feared ridicule, or, as he put it, he feared the clamor of the Boetians, a figurative reference to a dull-witted Greek tribe~\cite{Kline-02}.

In his research on the existence of a Non-Euclidean Geometry, Gauss figured out hypothetical ``worms'' that could live exclusively in a bi-dimensional surface, as other ``beings'' could be able to live in spaces of four or more dimensions~\cite{Silvester}. It is interesting to mention that, trying to verify his theory, Gauss and his assistants measured the angles of a triangle formed by the peaks of three mountains, Brocken, Hohehagen and Inselsberg, which belong to the Harz Mountais, in Germany. The distance between two of them were 69,85 and 197~km, respectively. The sum of the internal angles of this triangle was $180^\circ$ and 14'',85. This result frustrated Gauss since the error were within the errors associated to the instruments he used to measure the angles~\cite{Kline,Kaku}.

Independently of Gauss, the mathematicians, the Russian Nikolay Ivanovich Lobachevski (1793-1856) and the Hungarian J\'{a}nos Bolyai (1802-1860) (son of Wolfgang), in 1832, demonstrated the existence of triangles which sum of the internal angles are less than 180$^{\circ}$~\cite{Lobachevsky,Bolyai}.

The German mathematician Georg Friedrich Bernhard Riemann (1826-1866), after the presentation of his \textit{Doktoratsschrift}, in December 1851, in G\"{o}ttingen University, about the Fourier series and what is now know as Riemannian surfaces, started to prepare himself to become \textit{Privatdozent} of this same University. So, at the end of 1853, he presented his \textit{Habilitationsschrif} together with three topics for the \textit{Habilitationsvortrag}. For his surprise, Gauss choose the third topic entitled ``\"{U}ber die Hypothesen, welche der Geometrie zu Grunden liegen'' (``On the Hypothesis that are on the Base of Geometry''), where he demonstrated the existence of triangles of which the sum of its internal angles could be greater than 180$^{\circ}$. This topic was timidly presented by Riemann in June of 1854, but it provoked a deep impact on Gauss, because it was a concrete expression of his previous ideas about a Non-Euclidean Geometry (today, Riemannian Geometry) that he was afraid to publish, as previously mentioned. Riemann's metrical approach to Geometry and his interest in the problem of congruence also gave rise to another type of non-Euclidean geometry. We are talking about a new geometry that cames out not by the rejection of parallel axioms, but rather by its irregular curvature.

It is important to remember that those geometries, today generically know as Non-Euclidean Geometries~\cite{Beltrami,Loria} influenced physical thought in 19th century~\cite{Martins}. They are consequences of the observation that the relaxation of the Parallel Postulate could give rise to two new interpretations. One, in the Hyperbolic Geometry of Bolyai-Lobachevsky~\cite{Milnor}, for which, from a point outside a line, an infinite number of parallels can be drawn, and the second, in the Spherical Geometry of Riemann, where from a point outside a line, no parallel can be drawn to it~\cite{Kline,Bonola}.

\section{The popular interest in the new geometries}\label{popular}

Two specific areas of philosophical debate were the initial source of a \textit{sui generis} public interest in the non-Euclidean geometries and in the geometry in higher dimensions: the nature of the geometric axioms and the structure of our space~\cite{Henderson}. As time went on, a expressive interest of the general public fell on the nature of the space and the number of its dimensions. A historical record of this fact can be found in the accurate bibliography prepared by Duncan Sommerville (1879-1934), a Scottish mathematician and astronomer~\cite{Sommerville}. This history is well documented in the interesting book of Linda Henderson (b.~1948), historian of art~\cite{Henderson}. According to her, everything started with a movement to popularize $n$-dimensional spaces and non-Euclidean geometries in the second half of the 19th century. A whole literature was developed~\cite{Sommerville} around philosophical and mystical implications in relation to spaces of larger dimensions, easily accessible to a public of non-specialists; in particular, about the imagination of a fourth dimension, long before Minkowski's work and Einstein's Special Relativity and the Cubism. The popularization of these ideas contributed, as carefully analyzed in Ref.~\cite{Henderson}, to a revolution in Modern Art and, in particular, was fundamental to the Cubism, an artistic movement contemporary to Einstein's Special Relativity, where also use was made of non-Euclidean geometry, namely Minkowski's space-time.

\section{The fourth dimension as time-like component of the new space-time concept in Physics}
\label{sec:4}

It was Riemann who generalized the concept of Geometries, by introducing the definition of metric, that defines how one can calculate the distance between two points, given by (in nowadays notation)
$$\mbox{d} s^2 = \sum_{i,j} = g_{ij}  \mbox{d} x^i \mbox{d} x^j; \qquad (i,j = 1,2,3)$$
where $g_{ij}$  is the metric tensor of Riemann. In the case of flat spaces and rectilinear coordinates ($x,y,z$),
$$g_{ij} = (e_i, e_j) = \delta_{ij}$$
where $\delta_{ij}$ is the Kronecker delta,  $e_i\, (i=1,2,3)$ are the vector-basis of a particular coordinate system and the notation $(e_i, e_j)$ means the scalar product between the two vectors.

Thus, the distance can be written as
$$\mbox{d} s^2 = \sum_{i,j} = \delta_{ij}  \mbox{d} x^i \mbox{d} x^j\, = \, \mbox{d} x^2 + \mbox{d} y^2 + \mbox{d} z^2 $$
\noindent known as the Euclidean metric. This definition is straightforwardly extended to higher $n$-dimen\-sional spaces just doing $i,j \rightarrow \mu,\nu = 1,2,3, \cdots n$.

The Riemann work about Non-Euclidean Geometry (which easily allows the existence of more dimensions than the usual three), was soon recognized and flourish in all Europe, with eminent scientists propagating his ideas to the general public. For example, the German physicist and physiologist Hermann Ludwig Ferdinand von Helmholtz (1821-1894) considered Gauss’ worms leaving now in a Riemannian surface (on a sphere). However, in his book entitled \textit{Popular Lectures of Scientific Subjects}, published in 1881, he warned that it is impossible to represent (to visualize) the fourth dimension, because (...) such a representation is so impossible how it should be a color representation for someone born blind~\cite{Kaku-02}.

From now on, let us summarize the route of the assimilation of such ideas in Physics.

The success of Newtonian mechanicism will be put to the test, at first, by the study of heat made by the French mathematician and physicist Jean-Baptiste Joseph Fourier (1768-1830). In the Preliminary Speech of his \textit{Analytical Theory of Heat}, he states that
\begin{quotation}
\noindent\baselineskip=10pt {\small
\noindent ``Whatever the scope of mechanical theories, they do not apply to the effects of heat. These are a special type of phenomenon, and cannot be explained by the principles of movement and balance''.}
\end{quotation}

The propagation of heat will be described by a partial differential equation and no longer by an ordinary differential equation, as in the case of Newtonian mechanics. It is the beginning of valuing the \textit{causa formalis} over the \textit{causa efficiens} as the basis of the causal explanatory system in Physics, intrinsic to Newton's system~\cite{Causa_art}. It is the beginning of the description of Physics by Field Theories~\cite{Bachelard}. Later, in the second half of 19th century, also electromagnetism will reaffirm this trend~\cite{Oguri}. The discovery of electromagnetic waves by the German physicist Heinrich Rudolf Hertz (1857-1894) will give Maxwell's theory a new status. However, Maxwell theory is still a phenomenological theory not able to predict, for example, the interaction of light with matter. One of the first attempts to develop a classical interpretive theory capable of explaining the interactions of electromagnetic fields with matter dates from 1895 and is due to the Dutch physicist Hendrik Antoon Lorentz (1853-1928), who combines the Electromagnetism and Classical Mechanics with an atomistic model of matter, the so-called Drude-Lorentz model,\footnote{\,The model according to which the physical world would be composed of ponderable matter, electrically charged mobile particles and ether, such that electromagnetic and optical phenomena would be based on the position and movement of these particles.} and initially develops a Newtonian Classical Electrodynamics, known as Lorentz Electrodynamics.

Soon after the electron discovery, it gains a prominent place in theoretical physics~\cite{Buch}. In fact, as we have already mentioned, Lorentz will dedicate himself to include the interaction of this particle with the electromagnetic fields. As is well known, Lorentz Electrodynamics, inspite some  initial success, failed in correctly describe such kind of interaction. This problem will be solved just with the advent of Quantum Electrodynamics~\cite{Estranha}. From a conceptual point of view, Einstein attributed the weakness of Lorentz theory to the fact that it tried to determine the interaction phenomena by a combination of partial differential equations and total differential equations, a procedure that, in his opinion, is obviously not natural.

In 1888, the English Mathematician Oliver Heaviside (1850-1925) showed that the electric field ($\vec E$) of a moving electric charge (with velocity $v$) differ from that ($\vec E_\circ$) of a stationary charge as indicated below~\cite{Heav}:
$$\vec E_\circ = \frac{kq}{r^2}\, \hat r \quad \Rightarrow \quad \vec E = \frac{kq}{r^2}\,\gamma\, \left[\frac{1 - \beta^2}{1 - \beta^2 \sin^2\theta} \right]^{3/2}\, \hat r$$
where $\beta = v/c$ and $\gamma = (1-\beta^2)^{-1/2}$. So, we can see that, in the direction of motion ($\theta =0$), the electric field behaves like
$$\vec E_\parallel = \frac{1}{\gamma^2}\frac{kq}{r^2} \hat r$$
Therefore, this result was interpreted by Heaviside as a contraction of the electrostatic field.

This result was published in 1889 and it was discussed by Heaviside, the British physicist Oliver Lodge (1851-1940) and the Irish physicist George FitzGerald (1851-1901)~\cite{Gray}. Inspired on this result, FitzGerald proposed that the objects contract along their line of flight. Independently, Lorentz came to the same idea in 1892 (see footnote in Ref.~\cite{Lorentz-1895}). This is the origin of Lorentz-FitzGerald contraction, involving the $\gamma$ factor.

Pre-Minkowskian applications of non-Euclidean geometry in Physics weren't many and they were reviewed in Ref.~\cite{Walter}.

Now, we would like to stress that, although Lorentz demonstrated, in 1904, that time is related to tri-dimensional space through the relations known as Lorentz Transformations (LT)~\cite{Lorentz-orig}, it was only the Russian-German mathematician Hermann Minkowski (1864-1909) who understood~\cite{Minkowski} that the LT represent a kind of rotation in a 4-dimensional flat space having coordinates  ($x_1,x_2,x_3,x_4$), with a metric (measurement of the distance between two points in this space) defined by:

$$\mbox{d} s^2 =  \sum_{\mu,\nu}^4 g^{\mu \nu}  \mbox{d} x_\mu  \mbox{d} x_\nu  =  \, \mbox{d} x_1^2 + \mbox{d} x_2^2 + \mbox{d} x_3^2  +  \mbox{d} x_4^2$$

\noindent where $g^{\mu \nu} = \delta^{\mu \nu}$ is the four-dimensional Kronecker delta,  $x_1=x$, $x_2=y$, $x_3=z$, $x_4= ict$, and $i =\sqrt{-1}$.

This expression is known as the Minkowskian metric, or pseudo-Euclidean metric, due to the fact that it can be negative. Note that, to avoid the use of $\sqrt{-1}$, the mathematicians defined a signature for  $g_{\mu\nu}$, such that the indices $\mu$  and $\nu$  can assume the values 1, 2, 3, 4 $(+, +, +, -)$ with $x^4 = ct$, or 0, 1, 2, 3 $(+, -, -, -)$ with $x^0 = ct$, where $\pm$ means $\pm 1$ only on the main diagonal of the metric tensor~\cite{Caruso-02}.

In his seminal paper of 1905 about the Electrodynamics of Moving Bodies, Einstein were able to derive LT without having to resort to ether by postulating the constancy of light velocity in the vacuum, \textit{i.e.}, assuming it does not depend on the velocity of the moving body~\cite{Einstein,Miller}.

For Lorentz, the local time ($t^\prime$) introduced in the coordinate transformations between inertial references, would be just an auxiliary parameter necessary to maintain the invariance of the laws of Electromagnetism, as stated at the end of the second edition of his \textit{Theory of Electrons}~\cite{Lorentz}:

\begin{quotation}
\noindent\baselineskip=10pt {\small
\noindent ``If I had to write the last chapter now, I should certainly have given a more prominent place to Einstein's theory of relativity (...) by which the theory of electromagnetic phenomena in moving systems gains a simplicity that I had not been able to attain. The chief cause of my failure was my clinging to the idea that the variable $t$ alone can be considered as the true time and that my local time $t^\prime$ must be regarded as no more than an auxiliary mathematical quantity. In Einstein's theory, on the contrary, $t^\prime$ plays the same part as $t$; if we want to describe phenomena in terms of $x^\prime,y^\prime, z^\prime, t^\prime$ we must work with these variables exactly as we could do with $x,y,z,t$.''}
\end{quotation}

On the other hand, the conception and interpretation of Lorentz's transformations as a geometric transformation in a pseudo-Euclidean space of dimension 4, for Minkowskii, was only possible thanks to Einstein's assertion, as quoted in a meeting of scientists in 1908, in Cologne~\cite{Cologne}:
\begin{quotation}
\noindent\baselineskip=10pt {\small``But the credit of first recognizing clearly that the time of the one electron is just as good as the time of the other, that $t$ and $t^\prime$ are to be treated identically, belongs to A. Einstein.''}
\end{quotation}

However, another point made clear by Einstein is that for him the introduction of time as a fourth explicit coordinate in the transformations of inertial reference systems derives from the principle of relativity. In his words~\cite{Notas}:
\begin{quotation}
\noindent\baselineskip=10pt {\small``It is a widespread error that special theory of relativity is supposed to have, to a certain extent, first discovered, or at any rate, newly introduced, the four-dimensionality of the physical continuum. This, of course, is not the case. Classical mechanics, too, is based on the four-dimensional continuum of space and time. But in the four-dimensional continuum of classical physics the subspaces with constant time value have an absolute reality, independent of the choice of the reference system. Because of this [fact], the four-dimensional continuum falls naturally into a three-dimensional and a one-dimensional (time), so that the four-dimensional point of view does not force itself upon one as \textit{necessary}. The special theory of relativity, on the other hand, creates a formal dependence between the way in which the spatial coordinates, on the other hand, and the temporal coordinates, on the other, have to enter into natural laws.''}
\end{quotation}

\section{Concluding remarks}
\label{sec:5}

In this paper we have reviewed how mathematicians, physicists and philosophers have positioned themselves on whether or not a fourth dimension does exist. It was shown that the development of non-Euclidean geometries opened a new possibility to describe Physics. In addition to the classical example of Special Relativity, the possibility that extra spatial dimensions can play an important role in Physics is not new. It can be traced back to the pioneer works of Kaluza~\cite{Kaluza} and Klein~\cite{Klein}, in which a fifth dimension was considered, trying to unificate Electromagnetism and Gravitation. Following the general unification idea of Kaluza-Klein~\cite{Appel}, several higher-dimensional theories were developed, like String Theory and Supersymmetry~\cite{Schwarz,Dine}, based on theoretical ideas that go beyond the Standard Model of Particle Physics and show promise for unifying all forces. In all these examples, the extra dimension is always space-like.

Indeed, the introduction of extra dimensions in Fundamental Interactions Physics has been enabling a remarkable progress in two major contemporary programs: the quantization of gravity and the unification of the force fields of Nature, for which the mechanisms of reduction and dimensional compacting are of utmost importance~\cite{Helayel}.

It is interesting to point out that difficulties concerning the search for a Unified Theory of Elementary Physical Interactions (electromagnetic, strong, weak and gravitational) bring physicists to develop the \textit{M Theory}, which is a unifying theory in an eleven dimensional space (with just one temporal). Seven of those spatial dimensions are curled out and compactified in a Calabi-Yau space having dimensions equivalents to Planck's length  ($\approx 10^{-33}$~cm) and to them are attributed other properties, like mass and electric charge~\cite{Kaku}.

As a last remark, we can refer to the possibility of developing field theories with more than one coordinate time, in the course of 20th century, as reviewed in Ref.~\cite{Caruso-02}.

\end{document}